\documentclass[aps,prd,twocolumn,nofootinbib,amsmath,amssymb,showpacs]{revtex4-1}

\usepackage{graphics}
\usepackage{color}
\usepackage{hyperref}

\begin{document}
 
\title{Critical versus spurious fluctuations in the search for \\ the QCD critical point}

\author{M. {\sc Hippert}}
\email{hippert@if.ufrj.br}
\affiliation{Instituto de F\'\i sica, Universidade Federal do Rio de Janeiro,
Caixa Postal 68528, 21941-972, Rio de Janeiro, RJ, Brazil}
\author{E. S. {\sc Fraga}}
\email{fraga@if.ufrj.br}
\affiliation{Instituto de F\'\i sica, Universidade Federal do Rio de Janeiro,
Caixa Postal 68528, 21941-972, Rio de Janeiro, RJ, Brazil}
\author{E. M. {\sc Santos}}
\email{emoura@if.usp.br}
\affiliation{Instituto de F\'\i sica, Universidade de S\~ao Paulo,\\
Caixa Postal 66318, 05314-970, S\~ao Paulo, SP, Brazil}

\date{\today}

\begin{abstract}

The neighborhood of the QCD chiral critical point is characterized by intense fluctuations of the chiral field which 
could, in principle, generate pronounced experimental signatures. 
However, experimental uncertainties which are inherent to heavy-ion collisions, as well as 
the modest size and duration of the formed plasma, will severely attenuate these signatures.
Using Monte Carlo techniques, we study second-order event-by-event moments of  pions  
as a prototype for signatures of the chiral critical point based on the enhancement of the correlation length 
and event-by-event analysis.
We test their viability against some realistic ingredients, similar to the ones found in the RHIC Beam Energy Scan program.
\end{abstract}

\maketitle

\section{Introduction}
 
  The chiral phase diagram of QCD is believed to possess the very distinguishing feature of a second-order critical 
  end point \cite{Stephanov:2007fk,Rajagopal:1995bc,Stephanov:2004wx,PhysRevD.67.014028}. 
This point marks the end of a first-order transition line and its neighborhood exhibits very unique behavior. 
The discovery of the chiral critical point would represent a major breakthrough in the study of the phase diagram 
of strong interactions. 
Although many models agree on the existence of a critical point, there is no consensus about its location in the $(T,\mu_B)$ plane.
Furthermore, because of the sign problem, the reliability of lattice results is severely compromised for large values of the chemical potential. 
Nevertheless, current estimates strongly suggest the chiral critical point might be reachable at the current ultrarelativistic heavy-ion collision experiments (HICs), a possibility which 
brings the hope of accessing relevant information about the phase diagram of the strong interactions
\cite{PhysRevD.67.014028, PhysRevC.64.045202, PhysRevD.63.116009, PhysRevD.49.426, PhysRevD.58.096007}.

The existence, location and properties of the chiral critical point constitute a crucial issue in the study of the phase diagram 
of strong interactions.
Its clarification has been the subject of a large amount of theoretical work and is one of the main goals of the RHIC Beam Energy Scan program (BES) \cite{Aggarwal:2010cw}. 

One candidate source of experimental signatures of this point is 
the increase of long-wavelength fluctuations in its neighborhood, and, more specifically, 
its impact upon the event-by-event distribution of observables. 
Indeed, the use of event-by-event correlations of observables as signatures of the critical point in HICs was proposed and explored in the literature  
\cite{Stephanov:1999zu,Stephanov:2001zj,Stephanov:2008qz,Athanasiou:2010kw,Mukherjee:2015swa, Antoniou:2006mu, Antoniou:2007tr} and
 moments of the proton distribution are already being used in its experimental search \cite{Luo:2012kja, Adamczyk:2013dal, Luo:2013saa}.
Because of critical fluctuations, it is expected that these correlations should exhibit nonmonotonic behavior when the freeze-out conditions of the
 plasma formed in the experiments are varied across the neighborhood of the critical point, thus providing a possible signal of its presence.

However, while in equilibrium a second-order critical point is marked by extremely pronounced features, such as the divergence of susceptibilities, 
in realistic conditions equilibrium near this point is hardly attained and critical behavior can be 
dramatically attenuated by both dynamical and finite-size effects, especially in the specific context of the small, short-lived plasma formed in HICs 
\cite{Palhares:2010zz, Kiriyama:2006uh, Braun:2005fj, Berdnikov:1999ph, Stephanov:2009ra, Stephanov:2010zz, Mukherjee:2015swa, Antoniou:2006mu, Antoniou:2007tr}.
Moreover, since HICs are very complex experiments, it is not trivial that the nonmonotonic behavior emerging from critical fluctuations can 
be measured and the viability of such signals should be tested in realistic simulations.

Here, we use Monte Carlo techniques to test second-order moments of pions 
as signatures of the critical point in the context of HICs. 
While estimates of these signatures are available in the literature \cite{Stephanov:1999zu,Stephanov:2001zj,Stephanov:2008qz,Athanasiou:2010kw,Mukherjee:2015swa}, 
to our knowledge no previous attempt has been made to test them in a 
more realistic situation using computer simulations.

In order to introduce some realism, 
we discuss and include some sources of spurious contributions which are expected in these experiments.  
Nevertheless, we restrict our analysis to a very simple effective theory \cite{Stephanov:1999zu} and simplified models for both the spurious fluctuations and critical slowing down  \cite{Berdnikov:1999ph}.
Effects from dynamics are considered only in the estimates of the 
correlation length and finite-size effects are also partially taken into account by using Dirichlet boundary conditions on a sphere. 
We believe this simplified scenario to be optimistic when compared to the more complex situation of a real heavy-ion collision experiment, while taking into account 
some of the most essential features at play in such an environment.

\section{Fluctuations of pions}
\label{SecPionFluc}

\subsection{Effective theory}

The chiral field $\sigma$ provides an approximate order parameter 
for the chiral phase transition of QCD and is, for that reason, subject to long-wavelength fluctuations
 in the neighborhood of the critical point.  
Although this field is not directly observable, one would expect these fluctuations to have significant effect
upon pions, protons and essentially every field that interacts strongly enough with it, 
increasing event-by-event correlations among their corresponding observables \cite{Stephanov:2008qz,Athanasiou:2010kw}. 

We use the same effective theory as Refs. \cite{Stephanov:1999zu, Stephanov:2008qz}. Since we are interested in long-wavelength fluctuations, we adopt a classical approach along with a homogeneous approximation for the chiral field: $\sigma(\mathbf{x}) = \sigma_0$. 
One can then define a probability distribution for its zeroth mode, $\sigma_0$:
\begin{equation}
 P[\sigma_0] = e^{-\Omega[\sigma_0]/T}\,,
 \label{EqProbSigma}
\end{equation}
where the effective potential $\Omega$ can be expanded for small fluctuations of $\sigma_0$ (defining $\langle \sigma \rangle = 0$):
  \begin{equation}
   \Omega [\sigma_0] \approx V \, \dfrac{1}{2}\, m_\sigma^2 \; \sigma_0^2  + \mathcal{O}(\sigma_0^3)\, .
   \label{EffPot}
  \end{equation}

In Eq. (\ref{EffPot}), $m_\sigma$ is the physical mass of the chiral field, which goes to zero at the critical point.
Its value includes contributions due to interaction with other fields as well as thermal effects.

  We are interested in how the fluctuations of $\sigma_0$, given by Eq. (\ref{EqProbSigma}), affect fluctuations of 
  the pions. 
  For this purpose, it is appropriate to use the linear sigma model and take the coupling between the pions and $\sigma_0$ to lowest order in the fields, yielding the interaction 
  Lagrangian
  \begin{equation}
   \mathcal{L}_{int}  = - G\; \sigma_0 \, \vec \pi \cdot \vec \pi + \mathcal{O}(\phi^4)\, ,
\label{effCoupl}
  \end{equation}
which indicates that, in a first approximation, the interaction with $\sigma_0$ has the effect of correcting the pion mass:
\begin{equation}
 m_{\pi}^2 = m_{\pi}^{(0)\;2} + 2 \,G \; \sigma_0\, .
\label{pionmass}
\end{equation}
Expression (\ref{pionmass}), together with Eq. (\ref{EqProbSigma}), defines a Gaussian probability  distribution for 
the pion mass squared, of width $2 \,G \, \xi \,\sqrt{T/V}$, where the chiral correlation length $\xi$ is given by $m_\sigma^{-1}$. 
Hence one can calculate event-by-event moments of pions by considering an ensemble of realizations, each with a different pion mass, 
corresponding to chiral field fluctuations. This perspective is especially interesting for sampling Monte Carlo events.

Although Eq. (\ref{effCoupl}) is not manifestly invariant under chiral transformations, it should be clear that 
the employed coupling is derived from an approximately invariant potential, as explicit in Ref. \cite{Stephanov:1999zu}. 
We also note that the value of $G$ should depend on medium conditions, being estimated at $\sim 1900$ MeV in vacuum and $\sim 300$ MeV 
at the critical end point \cite{Stephanov:1999zu}, although we consider it to be constant in the vicinity of the critical point.

\subsection{Strategy}

In an ultrarelativistic heavy-ion collision with quark-gluon plasma formation, this plasma is expected to expand and cool. 
At a certain temperature the system undergoes chemical freeze-out, and inelastic collisions among the particles stop. 
After further expansion and cooling, kinetic freeze-out is achieved, and after some particles decay, a resulting distribution of 
particles is detected.
Repeating the experiment many times, one gathers a statistical distribution of observables, which contains information about the thermodynamic properties of the system at freeze-out and, 
hopefully, about whether the freeze-out conditions are near criticality.

We wish to reproduce critical contributions to second-order moments of pions in a Monte Carlo simulation. 
If we consider the system to be in equilibrium  and the only effect of interaction to be a correction $\delta m_\pi^2$ to the pion mass squared,  
the state of the pions at freeze-out is completely specified by the set of occupation numbers $\{ n_p \}$ where $p$ labels each eigenstate 
of the free one-particle Hamiltonian, 
including the momentum $\bf p$ and all relevant quantum numbers.
Thus, our strategy is to sample sets of occupation numbers for each event, corresponding to different values of $\sigma_0$ or, equivalently, $m_\pi^2$, 
 according to the effective theory outlined above. 
This can be accomplished by sampling a value for $m_\pi^2$ from Eqs. (\ref{EqProbSigma}) and (\ref{pionmass}), followed by each occupation number, sampled 
independently from the Boltzmann factor $e^{- \beta (\omega_p - \mu) \, n_p }$, where $\omega_p$ is the one-particle energy of the mode labeled by $p$ and 
depends on $m_\pi^2$.
In order to have a finite number of modes, it is necessary, of course, to work with a finite number of momentum modes, which can be done by 
imposing adequate boundary conditions and introducing a cutoff for the momentum.

It is possible to calculate critical contributions to second-order moments within this framework. 
Because of chiral field fluctuations, the correction to an energy level $\omega_p$, $\delta \omega_p = \delta m_\pi^2/2 \omega_p + \mathcal{O}(\delta m_\pi^4)$, fluctuates 
simultaneously for all modes, introducing correlated fluctuations among all occupation numbers.
This can be seen by calculating the microscopic correlator  $\langle \Delta n_p \; \Delta n_k \rangle$ with $p \neq k$, where $\Delta n_p$ denotes 
the fluctuation of $n_p$ about its average in a given event and  $\langle \cdots \rangle$ denotes an average over an infinite number of events,
\begin{multline}
 \langle  \Delta n_p \; \Delta n_k \rangle  = \dfrac{\mathcal Z_0}{\mathcal Z}\; \left\langle  \Delta n_p \; \Delta n_k \; \displaystyle 
 e^{-\beta \sum_{p^\prime} \delta \omega_{p^\prime}  \, n_{p^\prime}}  \right\rangle_0 \\
   = \dfrac{\beta^2}{8}  \;\langle (\delta m_\pi^2)^2 \rangle \;\displaystyle \sum_{p^\prime, k^\prime} \dfrac{1}{\omega_{p^\prime}} \dfrac{1}{\omega_{k^\prime}} \langle \Delta n_p \; 
\Delta n_k \; n_{p^\prime} \; n_{k^\prime} \rangle_0  + \mathcal{O}((\delta m_\pi^2)^4)\\
 \approx \dfrac{(G \xi)^2}{T V} \dfrac{1}{\omega_{p}} \dfrac{1}{\omega_{k}} f_p (1 + f_p)\, f_k (1+f_k) \, ,
\label{microC}
\end{multline}
which is in agreement with Ref. \cite{Stephanov:1999zu} and where we have used $\langle \Delta n_p \; \Delta n_k \rangle_0 = \delta_{p\,k} \; f_p (1+f_p)$, 
with the subscript $_0$ indicating nonperturbed averages, 
$\mathcal Z$ denoting the partition function, and $f_p =(e^{\beta(\omega_p - \mu)}-1)^{-1}$.

Near the critical point, the correlation length $\xi$ grows and Eq. (\ref{microC}) indicates that second-order moments of pions increase quadratically with it, 
providing possible signatures of criticality.
 In order to study higher-order cumulants, couplings of higher order in $\sigma_0$ should be included in Eq. (\ref{EffPot}). 
This is done in Refs. \cite{Stephanov:2008qz,Athanasiou:2010kw} and it is shown that these contributions are proportional to higher powers of $\xi$.

\section{Physical scenario}

The method presented in the previous section allows us to sample events that reproduce critical correlations among pions. 
However, this is clearly not enough to study the effect of these correlations in a realistic context 
--- useful analysis requires some detail on HICs and their relevant underlying background. 

It is hoped that the position of the critical end point could be revealed in these experiments through nonmonotonic behavior as the freeze-out conditions of the resulting plasma, 
namely its baryonic chemical potential $\mu_B$ and its temperature $T$, are varied around its neighborhood. 
Chemical freeze-out conditions should be more determinant for signatures involving particle multiplicities, while kinetic freeze-out conditions should 
affect signatures related to the transverse momenta spectra.
Nonetheless, these conditions are not directly controlled but rather estimated as functions of parameters such as the centrality class of the 
collision, its center-of-mass energy $\sqrt{s}$ and which ions are made to collide. 

Additionally, the nonmonotonic behavior we aim at might not be visible in experiments if it consists of a very small peak compared to background contributions. 
Its intensity will depend, among other things, on how much the chiral correlation length is allowed to grow when limited by finite-size and dynamical effects 
\cite{Palhares:2010zz, Kiriyama:2006uh, Braun:2005fj, Berdnikov:1999ph, Stephanov:2009ra, Stephanov:2010zz}.
The relevant background for our analysis comes from any source of noncritical fluctuations that affect the studied cumulants. 
This includes the variations of freeze-out temperature, chemical potential and plasma volume 
among collisions identified as freezing out in the same conditions, 
which might hide 
 thermodynamical event-by-event fluctuations. 
Dynamical effects are only roughly taken into account in the estimate of the maximum correlation length reached in the collisions, while finite-size effects 
are only partially included by boundary conditions which restrict the possible pionic modes.

 \subsection{Collision parameters}
 
 There are currently no precise predictions for the position of the critical point in the $(T, \mu_B)$ plane. 
A recent paper \cite{Lacey:2014wqa} claims to have found evidence of the critical point at $T \sim 165$ MeV and $\mu_B \sim 95$ MeV, although a 
lower limit to its baryonic chemical potential, $\mu_E \gtrsim 450$ MeV, was estimated in Ref. \cite{Fraga:2011hi} using finite-size scaling. 
This lower bound is consistent with lattice results excluding  $\mu_B \lesssim 500$ MeV and $\mu_B/T \lesssim 1$ for a chiral critical point \cite{Philipsen:2011zx,Moscicki:2009id}. 
Since we want to simulate realistic conditions, the idea is to take data from the RHIC Beam Energy Scan (BES) and simulate a situation 
in which the critical point is in reach for these experiments.

Since the considered thermal distribution of pions is not directly sensitive to the baryonic chemical potential $\mu_B$ in our treatment, the importance 
of experimental data is to provide a value of the freeze-out temperature, as well as to allow us to estimate the volume of the plasma, 
 as seen in Section \ref{GeoFluct}. 
We choose, for that purpose, data from STAR $Au + Au$ collisions at $\sqrt{s_{N N}} = 7.7$ GeV, for which $\mu_B = 420$ MeV at chemical freeze-out, 
which we consider to be sufficiently high \cite{Sandeep2014}.

For our simulations, we choose midrapidity 
and very high centrality, focusing at rapidity in the $-0.5<y<0.5$ range for the $5 \%$ most central collisions, 
with the purpose of reducing the effects of anisotropic flow and event-by-event fluctuations coming from 
initial conditions \cite{RafaelDerradi, Takahashi:2013gpa}.
Inspired by Refs. \cite{Sandeep2014,Adamczyk:2014mxp}, we use $T = 130$ MeV for the plasma temperature and $R_p = 6.8$ fm for its average radius 
in the $0\% - 5\%$ centrality class, standing midway between chemical ($T_{ch} = 145$ MeV, $R_{ch} = 5.8$ fm) and kinetic ($T_{kin} = 116$ MeV, $V_{kin} = 2000$ fm$^3$) 
freeze-out conditions. 
Since the strongest critical behavior is expected among the soft pions, and in order to avoid more complex behavior in the high transverse momentum region \cite{RafaelDerradi, Takahashi:2013gpa}, 
we consider only pions with transverse momentum $p_T$ below $1$ GeV.

\subsection{Critical slowing down}
\label{SecCritSlowDown}

One of the main limitations for the growth of the correlation length, and hence for the strength of signals of criticality, in HICs is the fact that, considering the finite duration 
of the formed plasma, there is not enough time for $\xi$ to grow by an arbitrarily large factor.
Thus, the very large increase in the equilibrium correlation length while approaching the critical point results in equally large equilibration times, 
 necessarily forcing the system out of equilibrium, a phenomenon known as critical slowing down \cite{Hohenberg:1977ym}. 

In order to estimate the highest attainable value of the correlation length in a collision, we inspire ourselves by Ref. \cite{Berdnikov:1999ph}, which 
takes advantage of static and dynamical  universality class arguments to make rough but robust estimates, largely based on qualitative behavior. 

The model of Ref. \cite{Berdnikov:1999ph} for the evolution of $\xi$ in time $t$ can be rewritten in the following way
\begin{equation}
 \dfrac{{d} \xi}{{  d} t} = A\; \left(\dfrac{\xi}{\xi_0} \right)^{2-z}\,\left(\dfrac{\xi_0}{\xi} - \dfrac{\xi_0}{\xi_{eq}(t)}\right)\, .
\label{modBerdnikov1}
\end{equation}
Eq. (\ref{modBerdnikov1}) is an educated guess based on the behavior for small deviations from equilibrium and depends on the universal exponents 
$\alpha = 0.11$, $\nu=0.63$ and $z=2 + \alpha/\nu$ \cite{ZinnJustin:1998ci, Guida:1996ep}, as well as the dimensionless, nonuniversal parameter $A$, 
which cannot be directly estimated.
The model assumes the system to be in equilibrium until it reaches, at temperature $T_0$, a correlation length $\xi_0$  
large enough that universality arguments apply.
 
The function $\xi_{eq}(t)$ describes the correlation length for a system in equilibrium at temperatures lower than $T_0$. 
 In order to find it, some simplifying hypotheses must be used along with universality. Namely, the plasma is taken to cool down at fixed baryonic chemical potential, 
and its trajectory in the phase diagram is supposed to map to the Ising model phase diagram so that it is perpendicular to its first-order 
transition line, with the Ising magnetic field $h$ being approximated as linear in the temperature $T$. 
The temperature is also taken to decrease at a constant rate, a simplification which can be improved on, 
 although with no significant gain in the results \cite{Berdnikov:1999ph}. We optimistically consider the system to cool through the critical point.
From the Ising model universality class, $\xi_{eq} (h) \propto |h|^{-\nu/\beta \delta}$, where $\beta=0.326$, $\delta=4.80$  \cite{ZinnJustin:1998ci, Guida:1996ep}, and, within 
these approximations,
\begin{equation}
 \xi_{eq}(t) = \xi_0 \left| \dfrac{t}{\tau} \right|^{-\nu/\beta \delta}
\label{modBerdnikov2}
 \end{equation}
where we denote by $\tau :=\frac{T_0-T_E}{|dT/dt|}$ the time interval taken by the system to cool down from $T_0$ to $T_E$ and choose $T(-\tau) = T_0$ and $T(0) = T_E$, 
where $T_E$ is the temperature at the critical end point.

\begin{figure}

\includegraphics{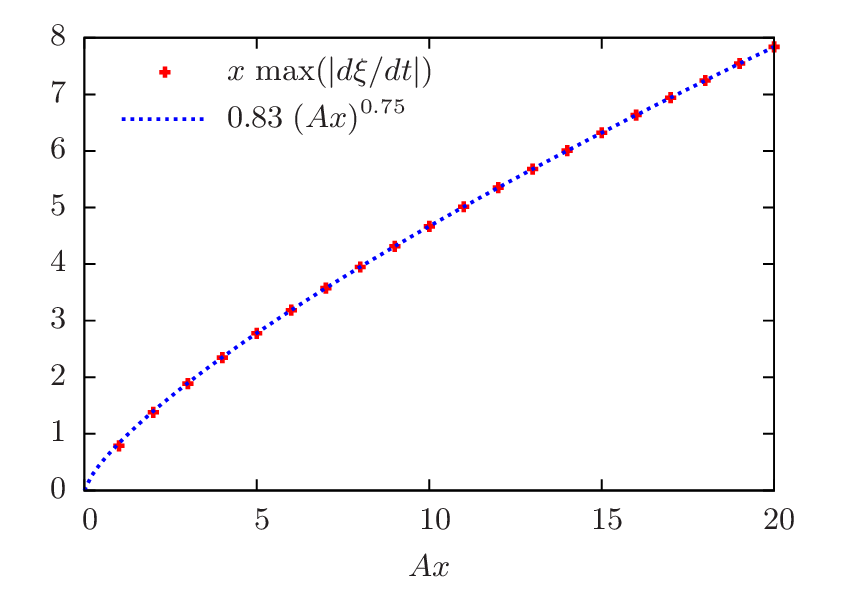}

	\caption{Relation between the unknown constant $A$ and the maximum value of $|d \xi/d t|$. Restricting the growth of $\xi$ to be below the speed of light 
	 yields the constraint $0.83 \; (A\, x)^{0.75} \leq x$.}
\label{DeriAmax}
\end{figure}

Ref. \cite{Berdnikov:1999ph} lacks an estimate for $A$, a very important parameter since it determines how closely $\xi$ follows its equilibrium value and, consequently, 
at how high a value it peaks. 
However, it is possible to put an upper bound on $A$ by requiring that $|d \xi/ d t|$ does not exceed the speed of light. 
The maximum value of $|d \xi/ d t|$ can be determined as a function of $A$ and the ratio $x:=\tau/\xi_0$. 
A plot of the product $x \;{\rm max} (|d \xi/ d t|)$ {\it versus} the combination $A\, x$ is shown in Fig. \ref{DeriAmax}.
The resulting curve is very well described by $x \;{\rm max} (|d \xi/ d t|) = 0.83 \; (A\, x)^{0.75}$, 
yielding an upper bound $A \leq A_{max} = 1.3 \; x^{0.33} $.

\begin{figure}

\includegraphics{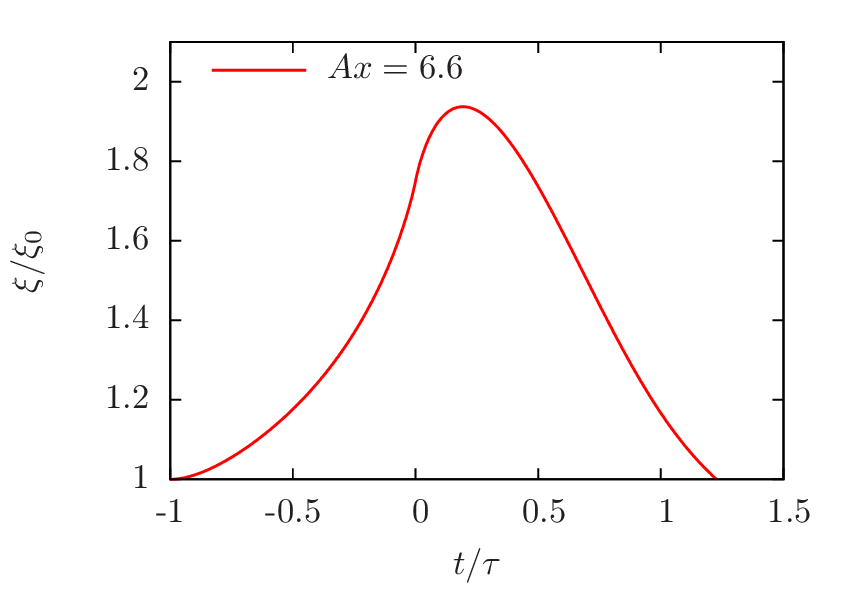}

	\caption{Evolution of the correlation length $\xi$ in the most favorable scenario, with $A=A_{max}$. 
	Its value never exceeds $2 \; \xi_0$, being most likely under $1.8\; \xi_0$, 
	depending on at which instant $t$ freeze-out occurs.}
\label{XiEvolution}
\end{figure}

The peak value of $\xi/\xi_0$ is controlled by the combination $A\, x$
\footnote{The dependence in the ratio $\tau/\xi_0$ can be easily interpreted, since a larger value of $\tau$ means a longer time reacting to criticality.}.
We choose $\tau = 5.5$ fm, inspired by Ref. \cite{Adamczyk:2014mxp}, 
 and, since $\xi_0 > 1/T_E$, take $\xi_0 = 1/120 \;{\rm MeV}= 1.6 \; {\rm fm}$, yielding 
  $x= 3.4$ and $A_{max} \; x= 6.6$.\footnote{In our case, $\tau = 5.5$ fm can be understood as taking $T_0 - T_E = 44$ MeV and $|dT/dt|= 8$ MeV/fm, although choosing the value of $\tau$ is less restrictive 
than fixing these quantities.}\textsuperscript{, }\footnote{The requirement 
$\xi_0>1/T_E$ comes from the conditions for the application of universality, namely $\xi\gg 1/T_E$, which is never exactly satisfied \cite{Berdnikov:1999ph}.} 
We consider these estimates to be quite optimistic, especially given that our value for $\tau$ is comparable to the lifetime of the system \cite{Adamczyk:2014mxp}. 
Fig. \ref{XiEvolution} shows the dependence of $\xi/\xi_0$ on $t/\tau$ 
on the most optimistic scenario, namely $A = A_{max}$. We note that, even in this scenario, $\xi$ can hardly exceed $1.8 \; \xi_0 = 2.9$ fm, reaching at most $2.0 \; \xi_0 = 3.2$ fm,  
its freeze-out value depending on how early freeze-out occurs.

\section{Spurious fluctuations}

Critical behavior is not the only source of event-by-event correlations in heavy-ion collisions. 
As the behavior one wishes to observe experimentally is marked by correlated fluctuations of observables 
of pions, any global fluctuation of experimental parameters is a source of background and should hence be 
taken into account. 
In fact, collisions at the same center-of-mass energy and in the same centrality class might correspond to slightly different 
parameters such as freeze-out temperature and plasma volume. 
These fluctuations of the parameters have a global effect upon the resulting particles which might be indistinguishable from critical collective behavior, 
affecting the measured correlations and providing background to critical fluctuations.

\subsection{Geometrical fluctuations}
\label{GeoFluct}

Information about the system size in heavy-ion collisions is usually obtained through centrality binning. 
This implies that events belonging to the same centrality class will in general have different volumes, with values within a given range. 
Volume fluctuations are expected to arise both from  a centrality bin width effect (CBWE), generated by variations of the volume within a centrality bin, and 
a centrality resolution effect, related to initial volume fluctuations \cite{Luo:2011ts, 0954-3899-40-10-105104, PhysRevC.87.044906}. 
For simplicity, we only consider CBWE volume fluctuations. 
In this case, it is possible to estimate the form of the volume distribution corresponding to a given centrality class by considering the probability distribution of 
values of the impact parameter. Furthermore, since the resulting distribution turns out not to be Gaussian, it is clear that volume fluctuations will also affect higher-order cumulants. 

From a geometrical argument, the number of ways in which two nuclei can collide with an impact parameter $b$ should be proportional to the perimeter $2\pi\, b$ 
of the circle  defined by it. Therefore, 
the corresponding probability is expected to be linear in $b$:
\begin{equation}
 \mathcal{P}(b) \propto b\;\;. 
\label{Probab-b}
\end{equation}
Also using simple geometry, and considering the colliding nuclei as discs,\footnote{Due to Lorentz contraction, treating the nuclei as discs is more reasonable than considering them to be spheres.}  
it is easy to calculate the transverse overlap area $A$ between two equal colliding nuclei as a function of $b$ and their radius $R_{N}$, 
\begin{equation}
 A(b, R_{N}) = 2 R_{N}^2 \cos^{-1}\left(\dfrac{b}{2 R_{N}} \right) - b \sqrt{R_{N}^2 - \dfrac{b^2}{4}}\, ,
\end{equation}
where $R_{N}$ can be taken from the parameter $r_0$ on the Woods-Saxon nuclear density 
profile,\footnote{In Refs. \cite{DeVries1987495,DeJager1974479} this density profile is referred to as the two-parameter Fermi model (2pF).} 
 $\rho (r) \propto (1 + e^{(r-r_0)/a})^{-1}$,
and we use the value $R_{N} = r_0 = 6.38$ fm (from Refs. \cite{DeVries1987495,DeJager1974479}).

Hence, one can get a probability distribution for the volume $V$ of the plasma at kinetic freeze-out 
by supposing that this volume is proportional to the initial overlap area $A$, with a proportionality factor $C$ with dimensions of length. 
In order to estimate $C$, we fix the average value of the plasma radius in the $0\% - 5 \%$ centrality class at $R_p = 6.8$ fm,  
yielding $C = 12.7$ fm \cite{Sandeep2014,Adamczyk:2014mxp}.

\subsection{Temperature fluctuations}

We know very little about the temperature  distribution for a given class of collisions and lack a simple model 
that connects geometrical and temperature fluctuations. 
For that reason, we just take a Gaussian distribution of temperatures, with a $5\%$ standard deviation. 

Even though the larger heat capacity near the critical point could diminish the impact of initial conditions on the freeze-out temperature, 
suppressing spurious temperature fluctuations, 
 its growth should also be limited by critical slowing down \cite{Hohenberg:1977ym}.
Since it should scale as $C_V\sim \xi^{\gamma/\nu}$ in equilibrium, where $\gamma = 1.240$  \cite{ZinnJustin:1998ci, Guida:1996ep}, 
 we just consider that scaling to hold as an approximation out of equilibrium and find out that $C_V$ can grow by at most $70\% - 130\%$ for a very limited interval of time, 
even considering sigma fluctuations to be responsible for $25\%$ of the heat capacity when $\xi=\xi_0$ and using $\xi/\xi_0 = 2 - 2.5$ \cite{Berdnikov:1999ph}.

\subsection{Additional sources of spurious fluctuations}

Until this point, we have only considered direct pions in our analysis. 
Although pions from resonance decays could provide a major source of background for 
the signatures we study here, our results already place a fairly stringent limit on their visibility \cite{Stephanov:1999zu}.
Hence, the results below should be regarded as the most optimistic scenario  while 
taking into account the minimal ingredients of an ultrarelativistic heavy-ion collision experiment. 
Experimental results, on the other hand, could be even less compelling. 
A study including pions from decays is left for the future. 

While a more refined analysis would also require taking longitudinal and transverse flow, as well as the corresponding event-by-event fluctuations, into account, 
the implementation of such effects in our Monte Carlo algorithm would imply a large demand for computing power and is also left for latter developments. 
This could be done using data from blast-wave model fits to experiments \cite{Abelev:2008ab,Hung:1997du,Das:2012yq, Schnedermann:1993ws}.

Since we are concerned with a limited window in phase space, we disregard effects from energy conservation.

\section{Results}

\begin{figure}
\includegraphics{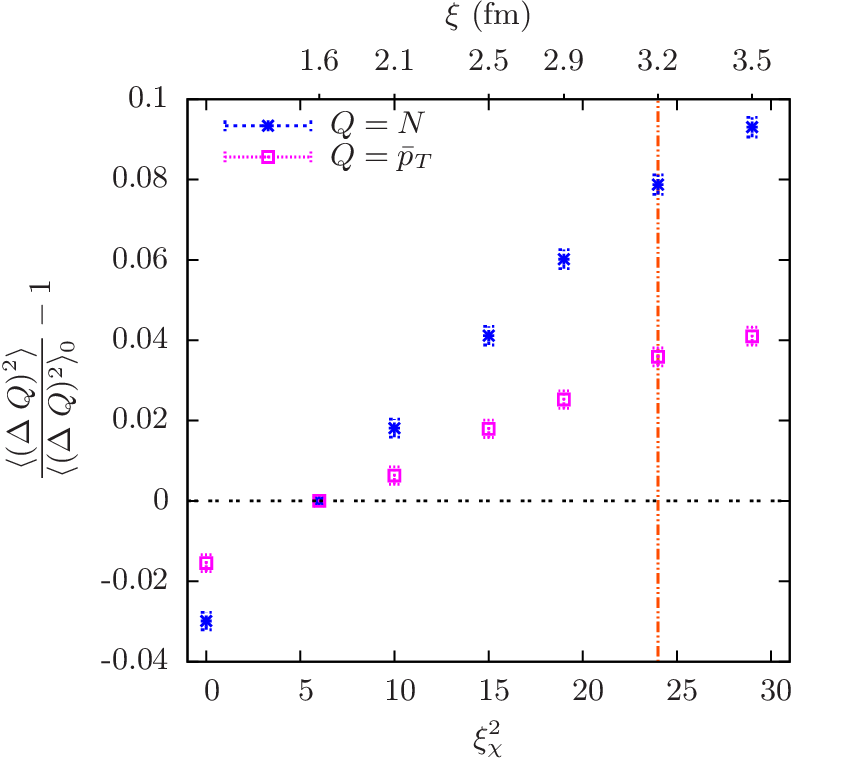}

	\caption{Signal as a function of $\xi_\chi^2$ and $\xi$, in proportion to the reference value, taken at $\xi_\chi^2= 6$, $\xi=1.6$ fm. 
The variances of the charged pion multiplicity, $N$, and the average transverse momentum of charged pions for a single event, $\bar p_T$, 
 are shown. The vertical dashed line, in red, marks the maximum value $\xi = 3.2$ fm found in Section \ref{SecCritSlowDown}.} 
\label{Result1}
\end{figure}

\begin{figure}

\includegraphics{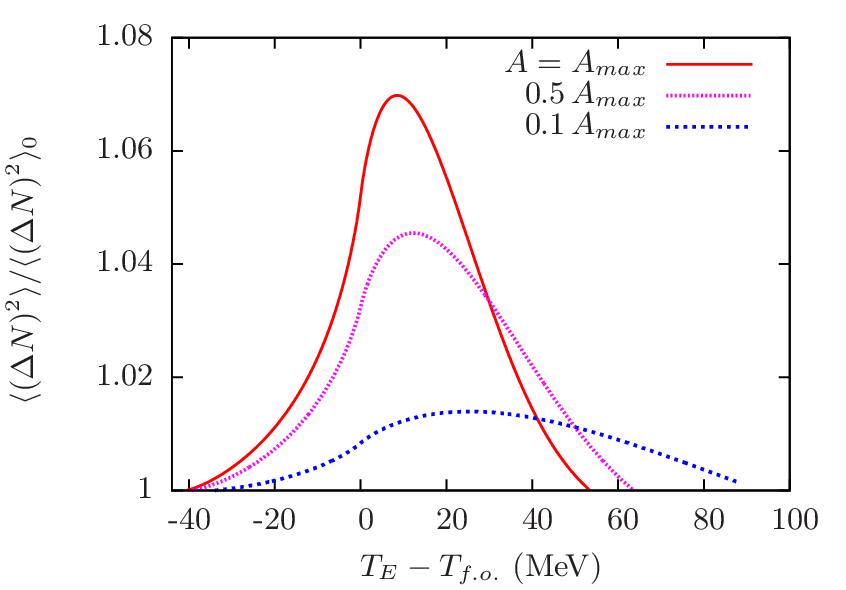}

	\caption{Behavior of the signal to baseline ratio of the variance of $N$ as the freeze-out temperature is varied for different choices 
of the parameter $A$.}
\label{SgnNNPlot}
\end{figure}

The model, methods and estimates discussed above 
were implemented in a Monte Carlo simulation, and samples of $10^6$ events were 
analyzed for several values of the chiral correlation length $\xi$ at  freeze-out. 
The viability of different  signatures of criticality could then be tested by comparing their values for increasing   $\xi$ and 
asking whether or not they are sufficiently distinct that some kind of nonmonotonic behavior 
might be detected while experimentally probing the critical end point neighborhood.

Since our model only includes $\xi$ through the combination $\xi_\chi^2 := (G \xi)^2$ and the estimated $G$ and $\xi$ are  
very uncertain, we display our results in terms of this quantity. A coupling $G = 300$ MeV is taken from Ref. \cite{Stephanov:1999zu}, 
where its value near the critical point is extracted from considerations using a Ginzburg-Landau effective potential and the linear sigma model. 
However, $G$ is expected to change significantly according to the medium, growing to a value of $G \approx 1900$ MeV in vacuum \cite{Stephanov:1999zu}, and 
it is difficult to estimate the value of $\xi_\chi^2$ far from the critical point. For that reason, we use 
$\xi_\chi^2 = (300 \;{\rm MeV} \cdot 1.6 \;{\rm fm})^2= 6$ as our baseline value.

Fig. \ref{Result1} exhibits our main result, consisting of the relative deviation from the baseline for the second-order moments 
$\langle (\Delta N)^2 \rangle$
 and $\langle (\Delta \bar p_T)^2 \rangle$ as functions of $\xi_\chi^2$, where 
$N$ is the charged pion multiplicity and $\bar p_T$ is the mean transverse momentum in a given event.
Both were normalized by the appropriate power of $\langle N \rangle$ in each sample in order to cancel out system-size dependence.
These signatures are linear in $\xi_\chi^2$, as should be anticipated from Eq. (\ref{microC}), and 
$\langle (\Delta N)^2 \rangle$ is observed to provide the strongest signal among them, reaching an increase of almost $10 \%$ depending on the value of $\xi$.\footnote{Results were succesfully compared with analytical estimates.}  
The quantity $F(p_T)$, suggested as a signature in Ref.  \cite{Stephanov:1999zu} was also calculated, showing increases of  $2.2\%$ for $\xi=2.9$ fm and 
$3.0\%$ for $\xi=3.2$ fm.

A higher signal, reaching up to $35\%$, is obtained by employing the mixed cumulant $\langle \Delta N \; \Delta \bar p_T \rangle$. 
However, given the temperature difference of $\sim 30$ MeV between chemical and kinetic freeze-out \cite{Sandeep2014}, 
it is very unlikely that a significant increase in $\xi$ due to criticality will affect both the particle multiplicity  and transverse momentum 
distributions. Hence, we discard this mixed signature as unrealistic.

By using a linear fit,\footnote{The mentioned linear fit yields good agreement, with a reduced $\chi^2$ of $2.3$.} 
it is possible to directly relate the signal in $\langle (\Delta N)^2 \rangle$ and the value of $\xi_\chi^2$, or, alternatively, 
$t/\tau$. 
Fig. \ref{SgnNNPlot} shows how the expected signal depends on the difference between the freeze-out temperature $T_{f.o.}$ and the critical point temperature $T_E$, 
assuming $T_0 - T_E = 44$ MeV and using different values for $A$. 
For a more moderate choice of $A$, such as $A_{max}/10$, the peak in this signature is decreased to a mere $1.4\%$. 

It is unclear whether the tested signatures could be visible in an experiment. 
Although we have found signals of almost $8\%$ for the maximum correlation length $\xi = 3.2$ fm, 
 our choice of parameters lies on the optimistic edge of their acceptable ranges ---  
we take the plasma to spend almost its entire lifetime near criticality and the correlation length to grow almost at the speed of light 
while following its equilibrium value.
We remark that changing the value of the time scale $\tau$  
has the same effect as changing $A$ by the same factor.

Additionally, albeit it contains some realistic features, we have explored a very simplified and optimistic scenario, 
assuming perfect equilibrium and neglecting factors such as
hydrodynamic flow and its related fluctuations as well as rescattering and resonance decay effects, 
all of which can have a strong impact upon the signal.
As an equilibrium distribution was considered for pions, even though $m_\pi^2$ fluctuates, 
we have implicitly neglected their equilibration times when compared to
 the time scale for chiral critical fluctuations. 
We take this as a simplifying hypothesis, albeit the equilibration time scales for the multiplicity and momenta of pions should further 
diminish the expected signatures, possibly completely ruining the signal in case these time scales are too large. 
One should also notice that 
 fluctuations of the freeze-out temperature and baryonic chemical potential should blur 
 the  sharp peak of Fig. \ref{SgnNNPlot} 
in case it is experimentally probed.
Bearing all these limitations in mind and using Fig. \ref{SgnNNPlot} as our reference, we find it rather 
unlikely that a signal, in case there is one, should go above $5 \%$ in a real experiment.

To gauge the importance of the spurious signal, the results discussed above were repeated without considering any effects from spurious fluctuations. 
It was found that in this case the variance of the multiplicity $N$ still provides the largest signal, although it is 
almost twice as sensitive to the value of $\xi_\chi^2$ and reaches almost $15\%$ for $\xi=3.2$ fm.
Our results indicate that spurious fluctuations only affect the signatures by dissolving the signal and  
simply  add a constant background contribution to the signatures, increasing the variance of $N$, normalized by its average, 
from $1.0$ to $2.0$ at $\xi = 1.6$ fm.
Regardless of that, this contribution is larger than the one coming from criticality and can definitely 
make the difference between a pronounced, detectable signature and one which is engulfed by noise. 

\section{Final remarks}
 
We have studied second-order event-by-event moments of pions as a prototype for  
signatures of the QCD critical end point based on the increase of the correlation length in its neighborhood. 
Those signals depend quadratically on the chiral correlation length 
and are therefore expected to exhibit nonmonotonic behavior as the neighborhood of the critical point is crossed by 
experimental conditions. In order to test this behavior against the experimental limitations of HICs, 
we have made use of simplified models and qualitative arguments, avoiding the introduction of unknown parameters 
and making estimates tending to be moderately optimistic.

We have found spurious fluctuations to have a significant impact upon the tested signatures.
The background in our simulations, provided by fluctuations of the freeze-out conditions, and
the limited growth of $\xi$, due to critical slowing down effects, were responsible for 
estimating the tested signals of criticality as probably less than $5\%$, even though the 
background was probably underestimated and our analysis is not particularly conservative. 

One way to make these signals slightly more pronounced would be to decrease contributions from geometrical 
fluctuations by using either a more restrictive centrality class or a centrality bin width weighting method \cite{0954-3899-40-10-105104,PhysRevC.87.044906}. 
However, in our simple model, geometrical fluctuations contribute only about $30\%$ of the background. 

An especially interesting extension to this work would be to test strongly intensive fluctuation measures, such as the $\varPhi$-measure for the transverse momentum 
\cite{Gazdzicki:1992ri}, as signatures of the critical point. 
These measures are constructed from extensive quantities to cancel out not only volume dependence but also 
effects from volume fluctuations and should thus display enhanced performance compared to ordinary 
cumulants \cite{PhysRevC.84.014904, *PhysRevC.88.024907, Gorenstein:2015ria}, although 
it is not clear for us how finite-size effects should affect them. 
However, the separation between kinetic and chemical freeze-out conditions makes our methods unreliable for 
combinations of quantities such as 
multiplicity and total transverse momentum. A relatively simple solution would be to use observables involving different particle 
species \cite{Mrowczynski199913, *Gazdzicki:1997gm}, 
but we leave this for future work.

Our results can also be generalized to second-order moments of protons, with the difference that the two flavors of charged 
pions would be replaced by the two spin states of the proton and quantum statistics would change.  
 However, while the pion mass is expected to be nearly the same at the critical point \cite{Stephanov:1999zu}, this is not the case for the mass 
of the proton, $m_P$, which is approximately proportional to the chiral condensate and should therefore 
be much smaller near the critical end point \cite{GellMann:1960np, Scavenius:2000qd}. 
The correction to the mass of the proton due to the chiral field fluctuations can be directly extracted from the 
linear sigma model and is $\delta m_P =g\, \sigma_0$, yielding $\delta (m_P^2) \approx G_P\, \sigma_0$, with $G_P = g\, m_P$ \cite{GellMann:1960np, Scavenius:2000qd}. 
In the vacuum, $f_\pi = 93$ MeV and $m_P^{vac} = 938$ MeV, resulting in $g= m_P^{vac}/f_\pi = 10$ \cite{Beringer:1900zz}. 
The rhs of Eq. (\ref{microC}) would then change by a factor of approximately $({G_P}/{G})\, ({p^2+m_\pi^{2}})/({p^2+m_P^{2}})$, 
which depends on how $g$ and $M_P$ depart from their vacuum values near the critical point. 
It is $\sim 1$ if $g$ preserves its vacuum value and $M_P \lesssim 30$ MeV near the critical 
point, so that protons are not expected to display much stronger signatures. 
Comparison between signatures from protons and pions can be found in Ref. \cite{Athanasiou:2010kw}.

Another possible generalization would be to include higher-order moments into our analysis, at the cost of introducing extra unreliable parameters in our model. 
These moments are expected to display stronger dependence on the correlation 
length  \cite{Stephanov:2008qz} and 
experimental results for their dependence on $\sqrt{s_{NN}}$ are available in Refs. \cite{Luo:2012kja, Adamczyk:2013dal, Luo:2013saa, Sahoo:2011at,Adamczyk:2014fia}.

\begin{acknowledgments}
We thank R. Derradi de Souza 
and M. Stephanov for helpful discussions. 
We are also grateful to P. Sorensen for significant comments. 
M. H. and E. S. F. acknowledge the kind hospitality of the ITP group at Frankfurt University, where part of this work was done.
This work was partially supported by CAPES, CNPq and FAPERJ. 

\end{acknowledgments}

\bibliography{Bibliografia}

\end{document}